\documentclass[a4paper,fleqn,review,aps,prb,superscriptaddress]{revtex4-2}
\usepackage{graphicx,float,amssymb,dcolumn,epstopdf}
\usepackage{subcaption}
\usepackage{CJK}
\usepackage[latin1]{inputenc}
\usepackage{tikz}
\usetikzlibrary{shapes,arrows}

\usetikzlibrary{automata}

\begin{document}
	\begin{CJK*}{GB}{}
		\title [mode = title]{Spin Hall Conductivity and Anomalous Hall Conductivity in Full Heusler compounds}
		
		\author{Yimin Ji}

		\affiliation{State Key Laboratory of Electronic Thin Films and Integrated Devices,
			University of Electronic Science and Technology of China, Chengdu, 610054, P. R. China}
		
		\author{Wenxu Zhang}
		\email{xwzhang@uestc.edu.cn}
		\affiliation{State Key Laboratory of Electronic Thin Films and Integrated Devices,
			University of Electronic Science and Technology of China, Chengdu, 610054, P. R. China}
		
		\author{Hongbin Zhang}
		\email{hzhang@tu-darmstadt.de}
		\affiliation{Institute of Materials Science, Technische Universt\"at Darmstadt, Darmstadt, 64287, Germany}
		
		\author{Wanli Zhang}

		\affiliation{State Key Laboratory of Electronic Thin Films and Integrated Devices,
			University of Electronic Science and Technology of China, Chengdu, 610054, P. R. China}
		
		\date{\today}
 	
\begin{abstract}
The spin Hall conductivity (SHC)  and anomalous Hall conductivity (AHC) in more than 120 full Heusler compounds are calculated using density functional theory in a high-throughtput way. The electronic structures are mapped to the Wannier basis and the linear response theory is used to get the conductivity. Our results show that the mechanism under the SHC or AHC cannot be simply related to the valence electron numbers or atomic weights, is related to the very details of the electronic structure, which can only be obtained by calculations. A high throughput calculation is efficient to screen out the desired materials. According to our present results, Cu$_2$CoSn, as well as Co$_2$MnAl and Co$_2$MnGa are candidates in spintronic materials regarding to their high SHC and AHC values, which can benefit the spin-torque-driven nanodevices.
\end{abstract}
\maketitle
\end{CJK*}
\section{Introduction}
Full Heusler compounds with chemical formula X$_2$YZ provide us an ideal platform to engineer electronic properties of the materials. The undistorted crystal structure (L2$_1$ or Cu$_2$MnAl type defined in the Pearson Table) is face-centered cubic with three elements occupying the three different Wyckoff positions: $8c$(X), $4a$(Y), and $4b$(Z) of the space group $Fm\bar{3}m$. The X and Y atoms are usually transition metals or lanthanides and the Z sub-lattice is always a main group metal or a semimetal. It has been shown that the compounds can be stable with varies combinations of X, Y and Z elements leading to tremendous varies of properties attached to them: half metallicity \cite{Rozale2013}, Weyl topological material\cite{Di2010}, thermoelectricity \cite{Fu2015,Zeier2016,Hinterleitner2019},\textit{et al}. The properties can be optimized to meet the demands in the areas. For example, the $z$T can be as huge as 1.5 \cite{Rogl2017} in thermoelectric application. The half metallicity in  Co$_2$YZ, for example, leads to the high magnetoresistance, which is crucial in its application in spintronic devices \cite{Elphick2021,Hirohata2020}. In the spintronic devices, the spin-orbital coupling (SOC) plays an important role. It is utilized to control spin textures, spin-charge conversions and spin precession etc. One of the important way to realized the spin-to-charge conversion is by the spin Hall effect (SHE), which converts a charge current into a spin current, or the inverse SHE (ISHE), which performances in the opposite way\cite{Sinova2015}. The key parameter the spin Hall conductivity (SHC) $\sigma_{\alpha\beta}^\gamma$ is written as $J_\alpha=\sigma_{\alpha\beta}^\gamma E_\beta$, where the spin is polarized in the $\gamma$-direction and the spin current $J$ is measured in the $\alpha$-direction when the electric field in the $\beta$-direction is  $E_\beta$. The SHE is either stemmed from spin dependent scattering of defects or the SOC dependent bands. The latter is a geometrical property of the bands which is named Berry curvature and is intrinsic in solids. Calculations of the SHC can be formulated from the linear response theory\cite{Guo2008,Guo2009} and the calculated values of Pt, Pd and Au etc.\cite{Guo2008,Guo2009} are in good agreement with the experiments. In the magnetic compounds, the spin to charge transportation conversion leads to the AHC. It was extensively investigated in the magnetic Heusler compounds due to its importance in spintronics. Early work by K\"ubler and Felser\cite{Kuebler2012} in the magnetically ordered compounds shows that the strength of SOC can enhance the anomalous Hall effect  under the condition that the other parameters are fixed. The increase comes largely from $X$ point in the BZ. However, when heavier elements are used, which on one hand enhances the SOC, on the other hand the chemical bindings are changed, as the values in Rh$_2$MnAl instead of Co$_2$MnAl is calculated, it does not show the expected increment. Giant anomalous Hall and Nernst effect in magnetic cubic Heusler compounds was investigated by Noky \textit{et al.} \cite{Noky2019,Noky2020}. A high-throughput calculations of the SHC in the 20,000 nonmagnetic compounds were recently reported by Zhang \textit{et al.} \cite{Zhang2021} and a database of the calculated SHC tensors is online. However, ferromagnetic order does not kill the effect. The ISHE in ferromagnetic alloy NiFe was reported by Chien \textit{et al.} \cite{Miao2013} about ten years ago. The spin Hall angle was estimated to be the same order as that of Pt. In addition to that, it provides us a way to tune the spin-to-charge conversion efficiency\cite{Kimata2019,Qu2020,Hibino2021} by varying the magnetization. Very recently, Leiva \textit{et al.} obtained a giant spin Hall angle about -0.19$\pm$0.04 in Weyl ferromagnetic Heusler compound Co$_2$MnGa, which is among the highest reported for a ferromagnet. Regarding to the rapid development of the field, systematic calculations of the SHC are in demand due to its importance in spintronic materials.

\par In this work, we calculated the SHC in full Heusler compounds from the ICSD database in a high-throughput way. We obtained the SHC in the nonmagnetic compounds and ferromagnetic ones as well. We analyze characteristics of the electronic structures and find that band mixing of the $t_{2g}$ manifold is a key ingredient to obtain large SHC values in these compounds.
\section{Calculation details}
The full Heusler compounds are crystallized in the face-centered-cubic (FCC) structure with space group of $Fm\bar{3}m$. The compounds were obtained from the ICSD database, which amounts to about 120. The lattice constant from the database is used in the calculation without optimization. The electronic structure calculations with the fully relativistic treatment of the kinetic energy were performed based on the full potential code (FPLO)\cite{fplo}. The electronic bands were then projected onto the Wannier functions with the minimum basis sets. Calculations of the spin Hall conductivity were performed by the code developed by ~{\v{Z}}elezn{\'{y}} \cite{shccode}, which evaluates the Kubo's formula within the linear response theory. The intrinsic SHC can be obtained by computing the spin Berry curvatures \cite{Guo2008}, which has the following form
\begin{equation}
\sigma^\gamma_{\alpha\beta}=\frac{e}{\hbar}\sum_n\int_{BZ}\frac{d^3\vec{k}}{(2\pi)^3}f_n(\vec{k})\Omega^\gamma_{n,\alpha\beta}(\vec{k}),
\end{equation}
where $\Omega^\gamma_{n,\alpha\beta}(\vec{k})$ is referred as the spin Berry curvature defined as
\begin{equation}
\Omega^\gamma_{n,\alpha\beta}(\vec{k})=2i\hbar^2\sum_{m\neq n}\frac{\langle u_n(\vec{k})|\hat{J}_\alpha^\gamma|u_m(\vec{k})\rangle \langle u_n(\vec{k})|\hat{v}_\beta|u_m(\vec{k})\rangle}{(\epsilon_n(\vec{k})-\epsilon_m(\vec{k}))^2},\label{equ:bc}
\end{equation}
with $\alpha,\beta,\gamma=x,y,z$ in the Cartesian coordinates, and $m,n$ being the band index. The spin current operator is $\hat J_\alpha^\gamma=\frac{1}{2}\{{\hat{v}_\alpha,\hat{\sigma}_\gamma}\}$, where $\hat{\sigma}_\gamma$ is the spin operator component $\gamma$, and $\hat{v}_\alpha$ is the velocity operator in $\alpha$-direction. The Fermi-Dirac distribution function $f_n({\vec{k}})$ is the mean occupation number of state $(n,\vec{k})$ at a finite temperature $T$. In this work, we set $T=300$ K.  The third-order tensor $\sigma^\gamma_{\alpha\beta}$ represents the spin current $J^\gamma$ generated by an electric field in the $\beta$-direction($E_\beta$) via $J^\gamma=\sigma^\gamma_{\alpha\beta} E_\beta$. The spin current is polarized in the $\gamma$ direction and flows in the $\alpha$-direction.
The AHC was calculated in a similar way, only the spin current operator $J_{s_\alpha^\gamma}$ is replaced by the velocity operator $v_{\alpha}$. In the zeroth order approximation, the AHE is just two copies of SHE in ferromagnetic metals, which was added up for the two spin channels. In the ferromagnets, the conductivity tensor is of four-order, We calculate the values where the magnetic moment is in the $z$-direction.
\par
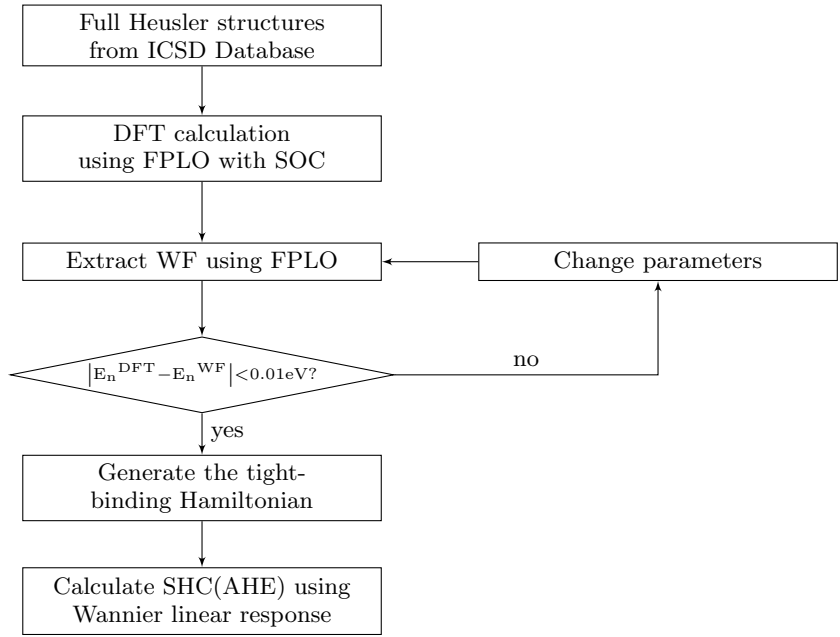
\begin{figure}
\centering
\tikzstyle{decision} = [diamond, draw,aspect = 5, text centered, inner sep=1pt]
\tikzstyle{line} = [draw,-latex']
\tikzstyle{box} = [rectangle,draw,text width=4.5cm,text centered]
\begin{tikzpicture}[node distance = 1.5cm,auto]
\node [box](first) {Full Heusler structures from ICSD Database};
\node [box, below of=first] (second) {DFT calculation \\ using FPLO with SOC};
\node [box, below of=second] (third) {Extract WF using FPLO};
\node [decision, below of=third] (fourth) {${\rm  \scriptstyle \left| {E_{n}}^{DFT}-{E_{n}}^{WF}\right|<0.01 eV?}$};
\node [box, right of=third,node distance=6cm] (third_r) {Change parameters};
\node [box, below of=fourth] (fifth) {Generate the tight-binding Hamiltonian};
\node [box,below of=fifth] (sixth) {Calculate SHC(AHE) using Wannier linear response};
\path [line] (first) -- (second);
\path [line] (second) -- (third);
\path [line] (third) -- (fourth);
\path [line] (fourth) -- node {yes}(fifth);
\path [line] (fifth) -- (sixth);
\path [line] (fourth) -| node[near start] {no}(third_r);
\path [line] (third_r) -- (third);
\end{tikzpicture}
	\caption{\label{fig:ht-process} Workflow for the high-throughput calculations of the SHE and AHE. All stable cubic full Heusler compounds from the database of Material Projects are taken as a starting input.}
\end{figure}

 The integral over the $k$-space during the calculation of the SHC was sampled in the full Brillouin zone with mesh grids of $200\times200\times200$ to ensure the convergence with respect to the $k$-mesh. The unit of SHC reported is $\hbar/e$($\Omega^{-1} $cm$^{-1}$). The whole process of the high-throughput calculation is shown in Fig. \ref{fig:ht-process}.

\par In the cubic crystal the nonzero elements can be written as $\sigma_{\alpha\beta}^\gamma=\sigma_0\epsilon_{\alpha\beta}^\gamma$, where $\epsilon_{\alpha\beta}^\gamma$ is the permutation symbol and $\sigma_0$ is a constant. It is anti-symmetric when $\alpha$ and $\beta$ are interchanged. When the compound is ferromagnetic, we fixed the spin polarization in the z-direction, and calculated $\sigma_{xy}^{z}$, which means we measure the voltages in the y-direction while the spin current is in the x-direction.
\section{Results and discussions}
\subsection{SHC and AHC around the Fermi level}
\begin{figure}
	\centering
	\begin{subfigure}{0.48\textwidth}
		\centering
		\includegraphics[width=\textwidth]{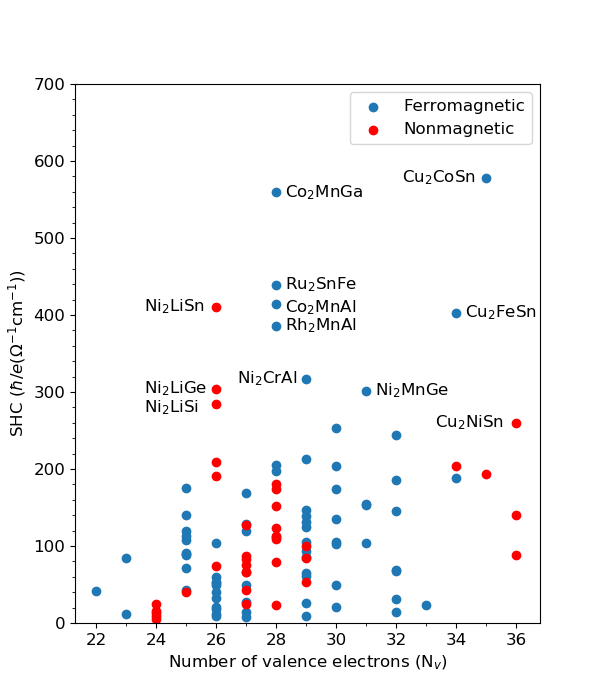}
		\caption{SHC at the Fermi energy}
		\label{fig:she-fermi}
	\end{subfigure}
	\hfill
	\begin{subfigure}{0.48\textwidth}
		\centering
		\includegraphics[width=\textwidth]{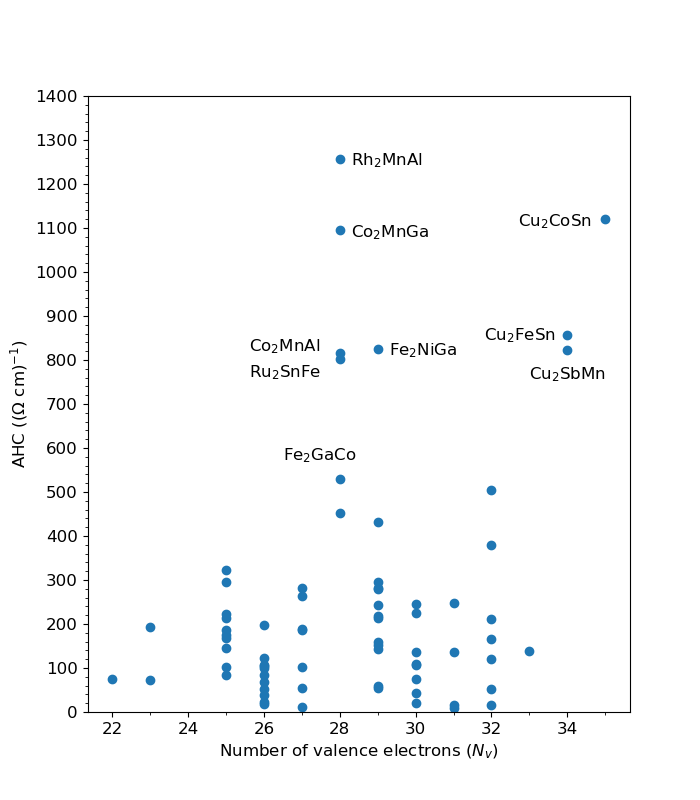}
		\caption{AHC at the Fermi energy}
		\label{fig:ahe-fermi}
	\end{subfigure}
	
	\caption{\label{fig:she-ahe-fermi} (color online) SHC and AHC  at the Fermi level of the Heusler compounds vs. the valence electron numbers ($N$). The ferromagnetic and nonmagnetic compounds are denoted with blue and red filled circles, respectively.}
\end{figure}

The non-zero SHC can be obtained for all the energy levels below and above the Fermi level normally. We can plot the SHC as a function of the Fermi level. However, only the electrons with energy around the Fermi level are relevant in most spin transport phenomena. We first show the values of $\sigma_{xy}^{z}$ at the Fermi level. They are  plotted with respect to the number of the valence electron as shown in Fig.\ref{fig:she-fermi}. The chemical formula of the compounds with large SHC are shown in the figure. All the data can be found in the Supplemented Materials.  We can see from the figure that the value reaches its maximum when the number of the valence electrons $N_v$ is 26. The largest value in the nonmagnetic compounds is about 400 $\hbar/e$($\Omega^{-1} $cm$^{-1}$) in Ni$_2$LiSn. In the ferromagnetic compounds, the maximum value is more than 500 $\hbar/e$($\Omega^{-1} $cm$^{-1}$) in Co$_2$MnGa and in Cu$_2$CoSn, with the magnetic moment about 4.1 $\mu_B$/f.u. and 1.1 $\mu_B$/f.u., respectively. The magnetic moment projected on Co and Mn is about 0.67 and 3.0 $\mu_B$ in Co$_2$MnGa, and on Co is 1.13$\mu_B$ in Cu$_2$CoSn. The non-integer magnetic moment indicates the partial filling of the d-orbitals and results in a low mobility of conduction electrons. The valence electron number dependent SHC can be partly understand from the electronic structure where the different orbitals are occupied as discussed in the next sections. From the data, we see that they are all below 2000  $\hbar/e$($\Omega^{-1} $cm$^{-1}$), which is the value typically obtained in Pt\cite{Guo2008}. We see that for the nonmagnetic compounds, for example, the SHC in Ni$_2$LiSn can reach as large as 400 $\hbar/e$($\Omega^{-1} $cm$^{-1}$) while in the ferromagnetic ordered Co$_2$MnGa and Cu$_2$CoSn, it is about 600 $\hbar/e$($\Omega^{-1} $cm$^{-1}$). The relatively small value of the SHC is rooted in the position of the Fermi level. It is not located at the place where the SOC mixes the states and opens a gap, which can enhance the Berry curvature. Thus the position of the Fermi level is crucial. However, in the experiments by Leiva \textit{et al}. \cite{PhysRevB.103.L041114}, the resistivity of Co$_2$MnGa thin film is about 220 $\mu\Omega^{-1}cm^{-1}$, which is ten times larger than that of Pt thin films.  When taken into account of this smaller electric conductivity in these compounds\cite{Elphick2021}, the spin Hall angle, which measures the efficiency of the spin to charge conversion efficiency, can be significantly larger than that in pure noble metals.
\par The AHC at the Fermi level is also shown in Fig.\ref{fig:ahe-fermi}. We can see that although the maximum of AHC obtained is in Rh$_2$MnAl. The same group of the compounds, Co$_2$MnGa, Co$_2$MnAl, and Rh$_2$MnAl can show significant high AHC and SHC among others with the same N$_v$. The overall distributions of both conductivities with respect to the number of valence electrons are similar. This similarity is rooted on the same mechanism of the both effects. In a simplified two band conducting model, if the spin-non-conserving contributions are overlooked \cite{Zhang2011,Tung2013}, $\sigma_{xy}^A=\sigma_{xy}^\uparrow+\sigma_{xy}^\downarrow$ and $\sigma_{xy}^S=(\sigma_{xy}^\uparrow-\sigma_{xy}^\downarrow)\frac{2\hbar}{e}$. The relative contributions of the spin channels can be measured by spin polarization $p^H$ of the Hall current written as
\begin{equation}
p^H=\frac{\sigma_{xy}^\uparrow-\sigma_{xy}^\downarrow}{\sigma_{xy}^\uparrow+\sigma_{xy}^\downarrow},
\end{equation}
where $\sigma_{xy}^\uparrow$ and $\sigma_{xy}^\downarrow$ are the spin-up and spin-down Hall conductivities, respectively. We listed the total magnetic moment, spin and Hall current polarization, SHC and AHC of the five compounds in Tab. \ref{tab:shc-ahc}. We read that except Rh$_2$MnAl, all others show absolute value of $p^D$ being around -1.0. The Co-based Heusler compounds were explored by Tung and Guo\cite{Tung2013}, in which Co$_2$MnAl and Co$_2$MnGa were highly proposed to be used in spintronics due to their simultaneously high AHC and SHC. According to our present results, Cu$_2$CoSn should be added to this material family of high Hall conductivity. We noticed the discrepancies of the values reported by the different works. It may come from the slightly different different lattice constants we used during the calculations. We uses the lattice constants determined by the experiments which also lead to the reduction of the magnetic moment.   

\begin{table}[b]
		\caption{\label{tab:shc-ahc}}{The magnetic moment (M),spin polarization ($p^S$), current polarization ($p^H$), SHC ($\sigma_{xy}^S$) and AHC ($\sigma_{xy}^A$) of five selected compounds which show relatively high SHC and AHC.}
		\begin{ruledtabular}
          \begin{tabular}{cccccccc}
			\textbf{Compounds} & $M$& $p^H$& $p^S$ & $\sigma_{xy}^S$&$\sigma_{xy}^A$ & $\sigma^{\uparrow}$ & $\sigma^{\downarrow}$ \\
			           &($\mu_B$/f.u.)&  &  &($\hbar/e$ $\Omega^{-1} $cm$^{-1}$)&($\Omega^{-1} $cm$^{-1}$)&($\Omega^{-1} $cm$^{-1}$)&($\Omega^{-1} $cm$^{-1}$)\\
			
			\hline
			Rh$_2$MnAl & 4.08&-0.61 & 0.74 & 386.3 & -1257.9 & -532.375 & -725.525  \\
			           & 4.1\footnotemark[1],4.06\footnotemark[2]&&&& -1723\footnotemark[1],1500\footnotemark[2]& & \\
			Cu$_2$CoSn & 1.09& 1.03  & -0.46 & 578.5 & 1119.5 & 704.375 & 415.125\\
			Co$_2$MnGa & 4.09&-1.02  & 0.63  & 559.6 & -1094.8 & -407.5 & -687.3\\
			           &4.11\footnotemark[1],4.128\footnotemark[3]&&&733\footnotemark[3]&-1310\footnotemark[1],1417\footnotemark[3]&&\\
			Co$_2$MnAl & 4.01&-1.01  & 0.67 & 414.9 & -816.8 & -304.67 & -512.12\\
			&4.04\footnotemark[1],4.037\footnotemark[3]&&&655\footnotemark[3]&-1631\footnotemark[1],1800\footnotemark[2],1265\footnotemark[3]&&\\
			Ru$_2$SnFe & 4.11&-1.10  & 0.70 & 438.9 & -801.3 & -290.925 & -510.375\\
		\end{tabular}
	\end{ruledtabular}
\footnotetext[1]{data from Ref.~\onlinecite{Noky2020}}
\footnotetext[2]{data from Ref.~\onlinecite{Kuebler2012}}
\footnotetext[3]{data from Ref.~\onlinecite{Tung2013}}
\end{table}

\par The data do not show any systematic variation with the atomic numbers, the atomic weights and the valence numbers which have been explored by us. This is due to the fact that the SHC are determined by the mixing of the states due to the SOC, which is dependent on the orbitals and the dispersions alike as shown in Equ.\ref{equ:bc}.


\subsection{The electronic structure of the compounds}
\begin{figure}
	\centering
	\includegraphics[scale=0.7]{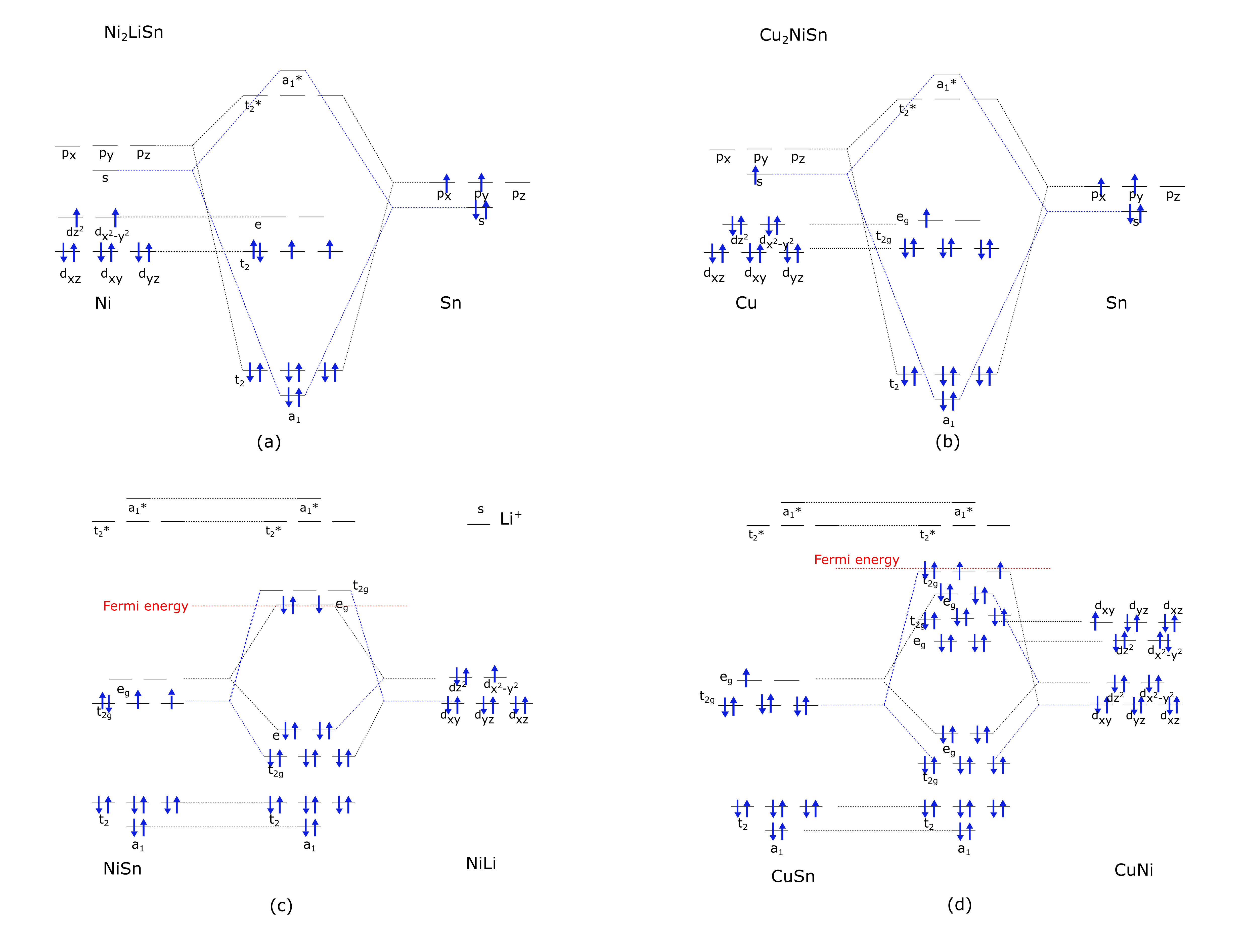}
	\caption{\label{fig:e_diagram} (color online) A molecular orbital analyses of the Heusler compounds. Different bondings of the electrons can be obtained due to the different electronegativity.}
\end{figure}

In order to get insight of the mechanism of the large SHC and AHC, we show the electronic structure of three selected compounds: Ni$_2$LiSn (N$_v$=26), Cu$_2$CoSn ($N_v$=35) and Co$_2$MnGa ($N_v$=28). These compounds have large SHC among the ones with the same number of valence electrons. The electronic structures of the Heusler compounds from the DFT calculations are explored in the published literatures. The basic characters of them can be understood from the molecular orbital model\cite{Graf2011}, including the crystal field splitting, the bonding of the atoms and the exchange splittings. For the nonmagnetic cases, it is schematically shown in Fig. \ref{fig:e_diagram}, in which we take Ni$_2$LiSn and Cu$_2$NiSn as examples. In a summary, we group the elements into X-Y and X-Z subgroups. For most compounds, where the X and Y element are from the TM group, the hybridization the sp-orbitals from X and main group element Z produces bonding and antibonding $a_1$ and $t_2$ representation of cubic crystal field as shown in Fig. \ref{fig:e_diagram} (a) and (b), while those of d-orbitals from X and Y form $t_{2g}$ and $e_g$ representations as in Fig. \ref{fig:e_diagram}(d). There are non-bonding $d$-orbitals. However, when Li instead of TM elements is at the Y site, ionic bonding are formed due to the large difference of the electron negativity of the elements. There are no additional nonbonding $d$-states. The cases $N_v=24$ and $N_v=35$ settle the Fermi level in the $e_g$ and $t_{2g}$ manifold respectively.

\begin{figure}
	\centering
	\begin{subfigure}{0.48\textwidth}
		\centering
		\includegraphics[width=\textwidth]{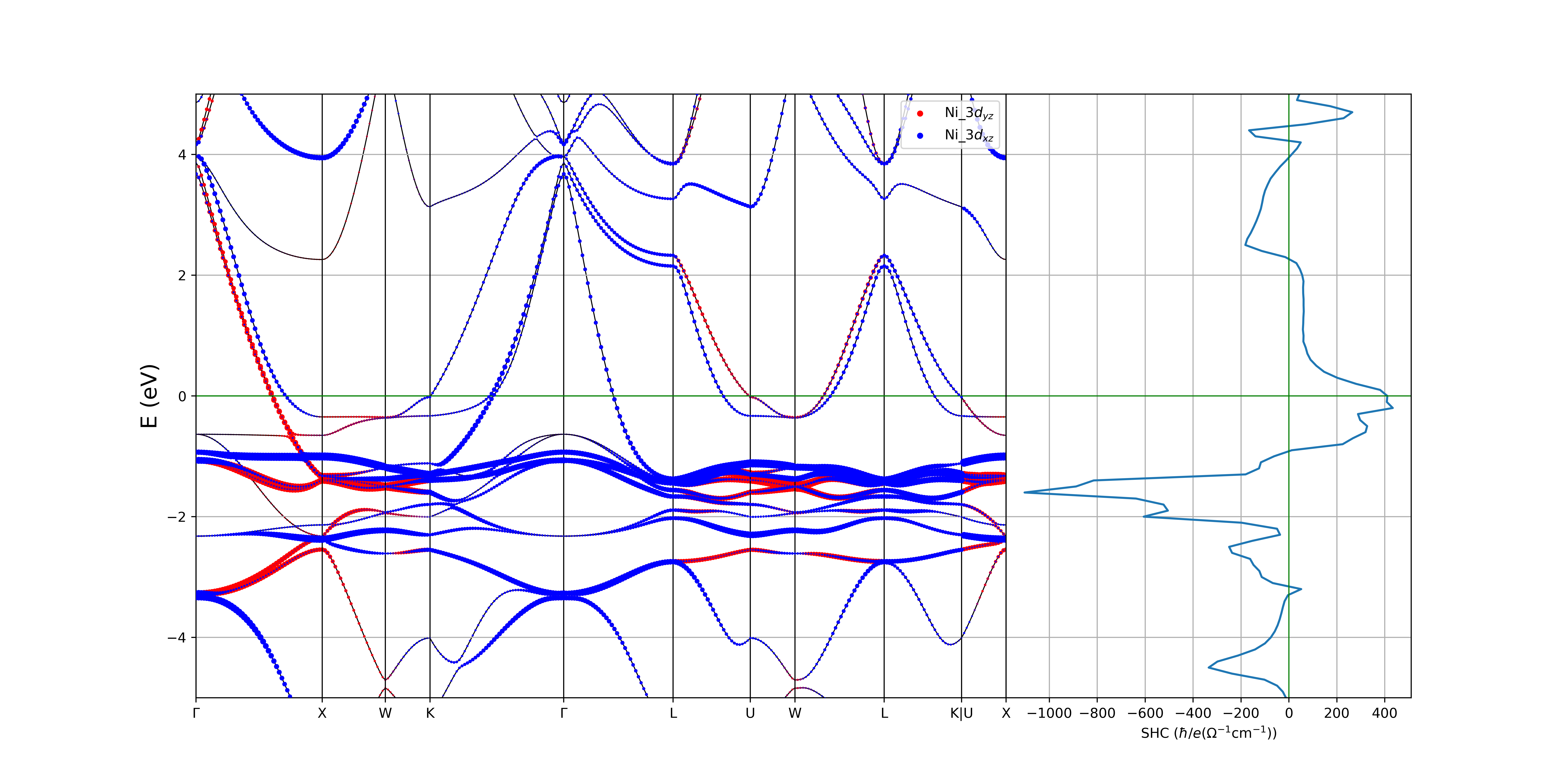}
		\caption{Ni$_2$LiSn}
		\label{fig:bands_Ni2LiSn}
	\end{subfigure}
\hfill
	\begin{subfigure}{0.48\textwidth}
	\centering
	\includegraphics[width=\textwidth]{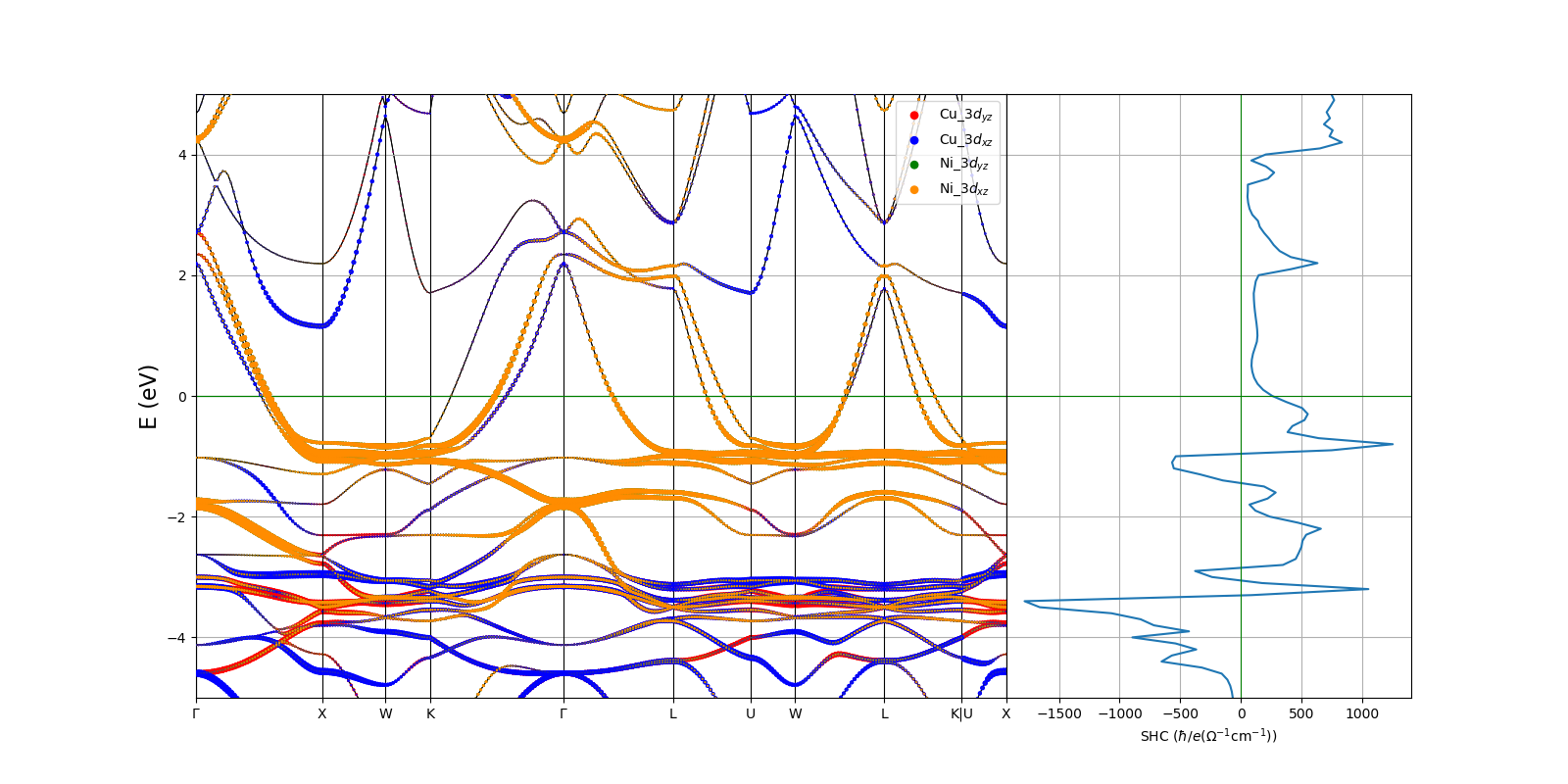}
	\caption{Cu$_2$NiSn}
 	\label{fig:bands_Cu2NiSn}
	\end{subfigure}
\hfill
	\begin{subfigure}{0.48\textwidth}
		\centering
		\includegraphics[width=\textwidth]{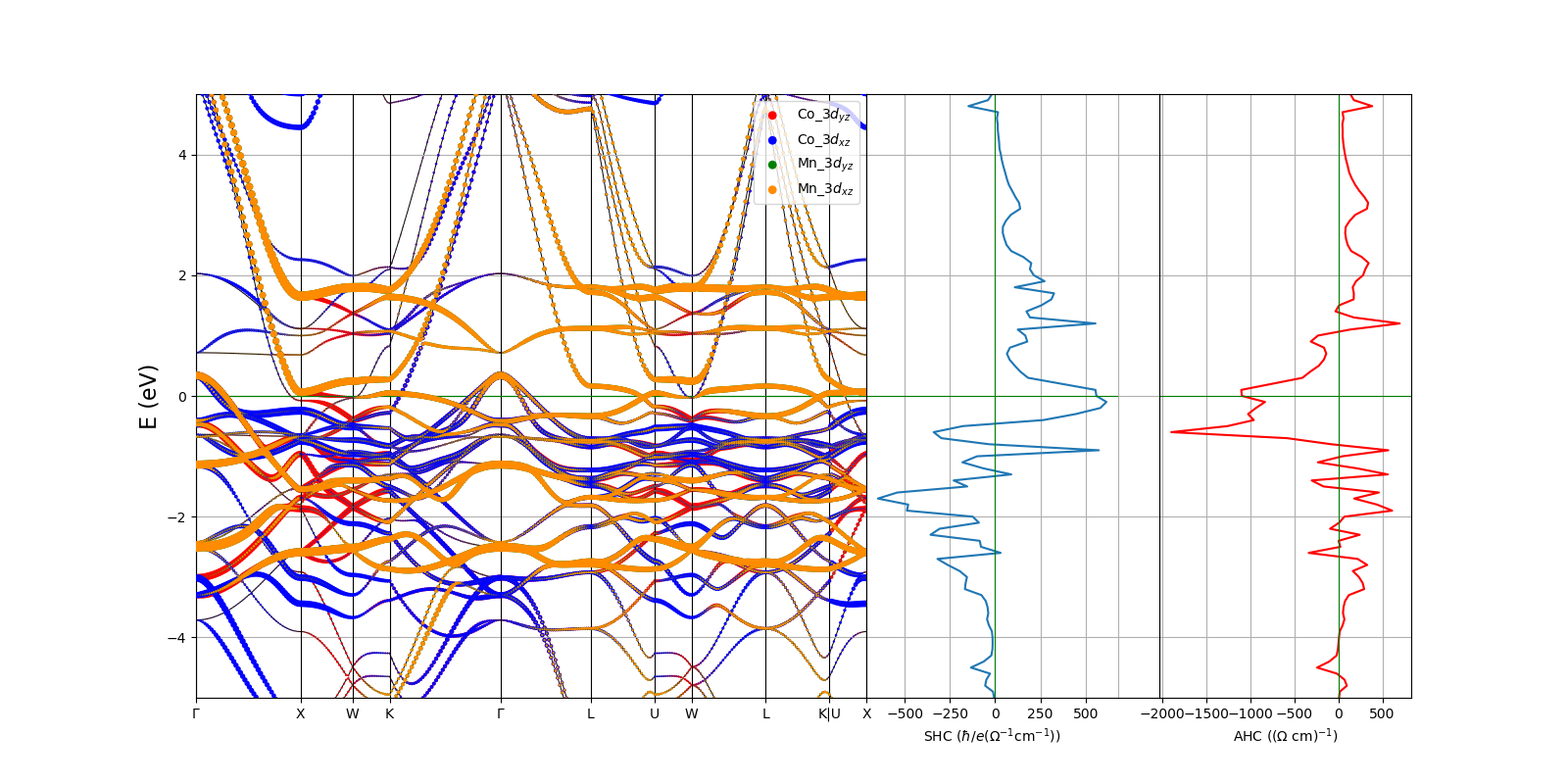}
		\caption{Co$_2$MnGa}
		\label{fig:bands_Co2GaMn}
	\end{subfigure}
\caption {The electronic band structures of three compounds with projected weights on $d_{xz}$ and $d_{yz}$ orbitals are shown. Each of the corresponding SHC is shown on the right panel. In ferromagnetic compound Co$_2$MnGa, the AHC is also shown as in (c). } \label{fig:bands}
\end{figure}

\par We show in Fig. \ref{fig:bands} the electronic bands of the three compounds with relatively high SHC from the \textit{ab initio} calculations. At the same time, we also show the weights from different atomic $d$-orbitals. In Ni$_2$LiSn, as shown in Fig. \ref{fig:bands_Ni2LiSn}, the Fermi level is among the $e_g$ states from Ni because of the crystal field as analyzed in the MO orbital model in Fig.\ref{fig:e_diagram}(c). The $t_{2g}$ states located mainly between -1.0 eV and -3.0 eV. When Y is the TM elements, as Cu$_2$NiSn and Co$_2$MnGa shown in Fig. \ref{fig:bands_Cu2NiSn} and \ref{fig:bands_Co2GaMn}, the $e_g$ and $t_{2g}$ states from the two subgroups form bonding and antibonding states. The nonbonding $d$-states in Cu$_2$NiSn are around -3 eV. The Fermi level crosses the antibonding states from Cu-Ni. It has the $t_{2g}$ characters. The exchange splitting makes the situation a little more complicated. It shifts the occupations of the states in Cu$_2$MnGa as shown in Fig. \ref{fig:bands_Co2GaMn}.
We see that in Cu$_2$MnGa the states near the Fermi level are coming from the 3d states of Mn. It is among the top of the t$_{2g}$ state of the majority spin and among the dip of the states of the minority spins.

\subsection{Hints for the large SHC and AHC values}
\par
\begin{figure}
	\centering
	\hfill
	\begin{subfigure}{0.48\textwidth}
		\centering
		\includegraphics[width=\textwidth]{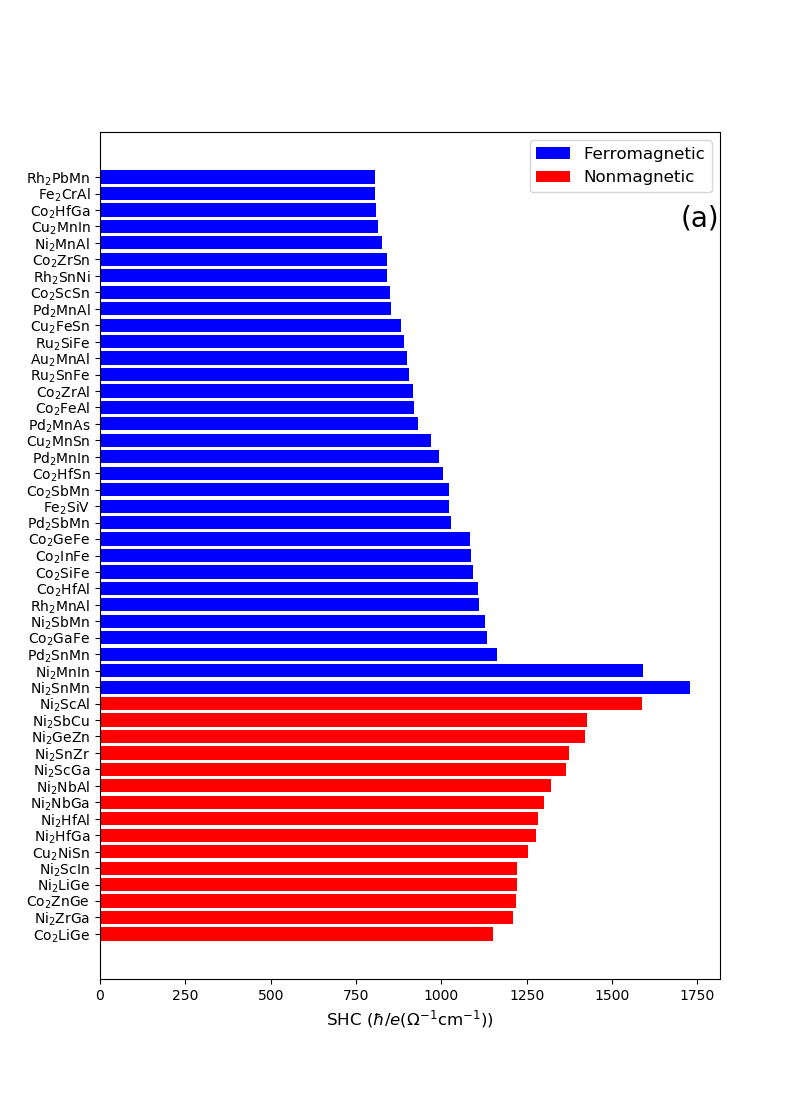}
		\label{fig:shc-max}
	\end{subfigure}
	\hfill
	\begin{subfigure}{0.48\textwidth}
		\centering
		\includegraphics[width=\textwidth]{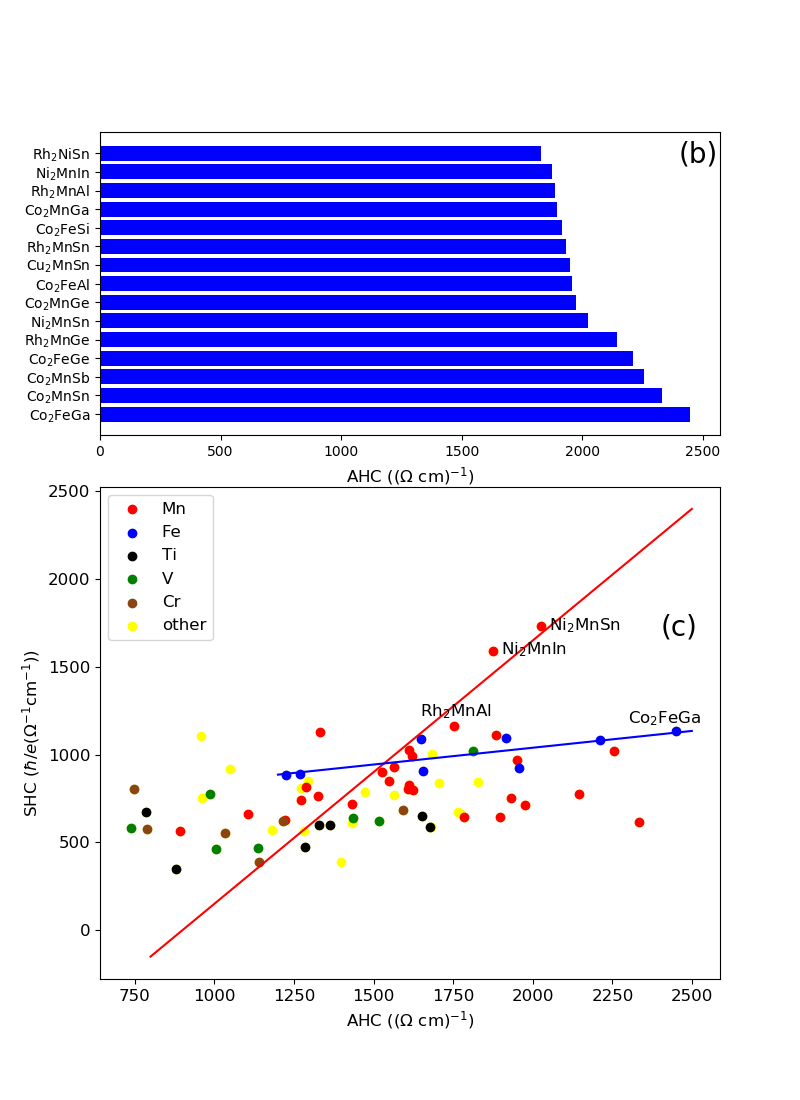}
		\label{fig:ahc-max}
	\end{subfigure}
	\caption{\label{fig:shc-ahc} (color online) The leading maximum of SHC (a) and AHC (b) of the compounds within $\pm2$eV around the Fermi level. The scattered values of AHC and SHC are shown in (c) with different colors representing different Y elements.}
\end{figure}

 The SHC is a function of the Fermi level. As can be seen in Fig.\ref{fig:bands_Ni2LiSn}, the maximum of the SHC in Ni$_2$LiSn is about -1000  $\hbar/e$($\Omega^{-1} $cm$^{-1}$) and locates about 1.8 eV below the Fermi level. We plot the maximum of the SHC absolute values within $\pm2.0$eV around the Fermi level achievable in the candidate compounds in Fig.\ref{fig:shc-ahc} (a). We see that in the figure the maximum is likely to be achieved in the magnetic compounds with Y = Mn and also in the nonmagnetic Ni$_2$YZ. According to our results in Fig. \ref{fig:shc-ahc}(a), the SHC in nonmagnetic compounds reaches its maximum about 1600  $\hbar/e$($\Omega^{-1} $cm$^{-1}$) in Ni$_2$SnCu. The values in Ni$_2$ScAl and Ni$_2$CuSb are a little bit smaller. We noticed that these compounds contains no elements such as Pt, W, or Bi, with large atomic SOC, However, the SHC is still comparable with that in materials contains heavy element. The value of AHC can exceed 2000 $\Omega^{-1}$cm$^{-1}$ in Co$_2$MnSb(Sn) and Rh$_2$MnGe as shown in \ref{fig:shc-ahc}(b). Both values are plotted as scattered circles filled with different colors for different Y elements as shown in \ref{fig:shc-ahc}(c). Generally, a larger AHC also leads to a larger SHC. We fitted the data for Y = Fe and Mn respectively. We noticed that when Y = Fe, the data are less scattered, because of the near half-metal characteristics observed in the DOS. This will minimize the spin-non-conserving contributions.  For the other groups of the compounds, the data are too scattered. The straight line fitting is given just for guiding sight.       

\par It was argued by Banerjee \textit{et al.} that the mixing of the pairs of \{$d_{xz},d_{yz}$\} or \{$d_{xy},d_{x^2-y^2}$\} orbitals due to the SOC is the main underlying mechanism for the anomalous velocity in TMO\cite{Jadaun2020}. $d_{xy}$ and $d_{x^2-y^2}$ orbitals belong to the different representation basis of the cubic group. According to the molecular orbital picture of the Heusler compounds, the hybridization of the X and Y orbitals and then to the orbitals of $Z$ and the second $X$ orbitals leads to the formation of  $a_1$-$t_2$, $e$-$t_{2g}$ manifold due to the cubic symmetry of the crystal field. The hybridization of the orbitals with $m=\pm 1$ is possible only within the $t_{2g}$ symmetry. It is anticipated that the Fermi level is located among the $t_{2g}$ manifold in order to obtain a larger effect. Accordingly, for the nonmagnetic compounds, significant SHC can be found in the compounds when the Fermi level crosses the $t_2$ states or the region when $e_g$ and $t_{2g}$ manifold overlaps with $N_v$ less than 24, or larger than 32. Unfortunately, in quite some Heusler compounds, the Fermi level is located in the gaps formed due to the crystal field\cite{Graf2011}. It is not expected that the Fermi level crosses the region of the band mixing, so that the SHC is not large. It is very occasionally that the Fermi level is in the vicinity where the SOC mixes the states and opens a gap. We may conclude that tuning the position of the Fermi level is the key to tune the SHC. Although the SHC at the Fermi level is not large, as we argued in the above section, it is due to the position of the Fermi level. When we change the Fermi level, the SHC can be significantly enhanced as shown in Fig. \ref{fig:bands_Ni2LiSn}. In the nonmagnetic states, the $t_{2g}$ states in Ni$_2$LiSn from Ni $d_{yz}$ and $d_{xz}$ orbitals are located mainly in the energy range between -1.0 eV and -2.0 eV. It shows very pronounced SHC in the energy window. Hybridizations of the two orbitals occurs at the W-point, which is in agreement with scenarios proposed by Jadaun \textit{et al}.\cite{Jadaun2020} Therefore, as already commented by K\"ubler \emph{et al.}, the AHC does not only determined by the strength of atomic SOC, but also related to the details of the electronic structure, namely, the interactions of the atomic orbitals. From the data obtained so far, we cannot draw any conclusions of the principles to obtain large SHE in Heusler compounds, although the nodal lines are proposed as a character of the large SHC/AHC\cite{Zhang2021}. Actually, the values depend on the electronic band structures, and the wave functions as well. We can only figure out the places in the band structures responsible for the SHC/AHC peaks\cite{Tung2013}. Thus a high-throughput calculation is a necessary and efficient way to figure out compounds with large SHC\cite{Zhang2021}.

\section{Conclusions}
We calculated the SHC and AHC in 120 full Heusler compounds, both in nonmagnetic and ferromagnetic states. The largest SHC at the Fermi level of nonmagnetic compounds is about 400  $\hbar/e$($\Omega^{-1} $cm$^{-1}$) in Ni$_2$LiSn and about 550 $\hbar/e$($\Omega^{-1} $cm$^{-1}$) in ferromagnetic Co$_2$MnGa. The largest AHC is about  We find that the electronic structure can be well understood by the MO models, where electronic states formed by the $t_{2g}$ and $e_g$ representations of the cubic symmetry play the determinant role of properties. However, the SHC is dependent on the subtle of the electronic structure. The wave function plays a role where the mixing of the states due to SOC enhanced the values. The Heusler compounds, due to their varieties of properties, may find more applications in modern electronic devices and systems.

\begin{acknowledgments}
Discussions with Jakub \v{Z}elezn\'{y} were greatly acknowledged. W. Z. acknowledged the financial support from the National Key R\&D Program of China (No.2017YFB0406403).
\end{acknowledgments}

\end{document}